\documentclass[10pt, a4paper]{article}
\usepackage[utf8]{inputenc}
\usepackage{graphicx}
\usepackage{amsmath}
\usepackage{amssymb}
\usepackage{geometry}
\usepackage{caption}
\usepackage{cite}
\usepackage{color}
\usepackage{float}

\title{\textbf{Design and Numerical Simulation of a SARA-Based RF Accelerator Using the $\mathrm{TE}_{112}$ Mode}}
\author{Tomás A. Carreño, Jesús E. López, Yerson F. Barragan, Carlos J. Páez-González,\\ Eduardo A. Orozco}
\date{}

\begin{document}
\maketitle

\begin{abstract}

\noindent Charged particle accelerators play a pivotal role in scientific research, industry, and medical applications. Among them, radiofrequency (RF) accelerators offer a promising approach for achieving high-energy particle acceleration in compact systems. This study presents the design and numerical simulation of a microwave-driven electron accelerator based on the Spatial Autoresonant Acceleration (SARA) mechanism. The proposed system consists of a cylindrical resonant cavity excited in the TE$_{112}$ mode, influenced by a tailored magnetostatic field generated by a set of external coils. The electromagnetic field distribution, magnetostatic configuration, and electron dynamics were simulated using COMSOL Multiphysics. The results validate the feasibility of accelerating electrons to energies close to 200~keV with a 1~kW microwave source, demonstrating their capability for X-ray generation through Bremsstrahlung radiation. Additionally, the study provides optimized design parameters for practical implementation. The high-quality factor $(Q\sim25000)$ of the resonant cavity ensures efficient energy storage but necessitates precise frequency control to maintain resonance conditions. The findings underscore the potential of SARA-based accelerators for compact X-ray sources, with applications in medical imaging, security screening, and materials analysis.

\end{abstract}

\section{Introduction}
 
\noindent Charged particle accelerators have been indispensable tools across various fields, ranging from fundamental physics to industrial and medical applications. These devices have played a crucial role in groundbreaking discoveries, such as the identification of quarks in the 1970s and the observation of the Higgs boson at the Large Hadron Collider (LHC) in 2012 \cite{discoveryquarks, Higgshistorical}. Such advancements have significantly deepened our understanding of the fundamental structure of matter and the interactions governing the universe. Beyond fundamental research, accelerators are widely used in industrial applications, including material sterilization and coating, radioisotope production for nuclear energy, and medical technologies such as cancer radiotherapy and X-ray imaging \cite{linacIndustry,therapyproton,therapyprotonphysics, historyxray}.

\noindent Given the broad relevance of these devices, research efforts have primarily focused on two major challenges: (i) Achieving higher particle energies to support fundamental physics investigations. (ii) Reducing construction costs to facilitate broader practical applications, particularly in the healthcare sector. In particular, compact and portable accelerators for X-ray generation are of great interest, as they would enable diagnostics in remote locations and emergency scenarios. While conventional acceleration schemes—such as linear accelerators (LINACs), cyclotrons, and synchrotrons—dominate the field, alternative approaches are being explored to address these challenges \cite{geng2024compact,suzuki2009development}.

\noindent One such alternative is electron cyclotron autoresonance acceleration, initially proposed by A. Kolomenskii and A. Levedev in 1962 and independently by V. Davydovskii in 1963 \cite{SelfResonanceacceleration,Resonanceacceleration}. This mechanism exploits the resonant interaction between an electron and an electromagnetic wave in the presence of an external static magnetic field. ECR acceleration enables electrons to reach X-ray-generating energies over short distances, sparking numerous theoretical and experimental studies aimed at optimizing the resonance conditions \cite{artarc1, artarc2, artarc3, artarc4}. Building upon this concept, K. S. Golovanivsky introduced the Gyro-Resonant Accelerator (GYRAC) mechanism \cite{1980PhyS...22..126G, gyracautoresonace}, which accelerates electrons by sustaining ECR conditions within a magnetic mirror trap with a time-varying magnetic field. The GYRAC concept was experimentally validated in 2017, achieving electron energies on the order of MeV \cite{Aplicationgyracautoresonace}.

\noindent Additionally, A. Neishtadt and A. Timofeev demonstrated that autoresonant acceleration plays a crucial role in electron cyclotron heating of plasmas. This occurs as electrons can maintain the ECR condition while moving along magnetic field lines toward regions of increasing magnetic field strength \cite{neishtadt1987autoresonance}. This mechanism, known to as spatial autoresonance, has laid the foundation for the design of accelerator cavities dedicated to electron beams \cite{PhySARA,dugar2017compact,otero2019numerical,velazco2003development,velazco2016novel}. In such systems, the magnetostatic field profile is carefully designed to preserve the resonance condition, ensuring that the microwave frequency matches the cyclotron frequency along the electron trajectory. A novel compact accelerator, known as the Rotating-Wave Accelerator (RWA), was proposed by Velazco et al  \cite{velazco2003development,velazco2016novel}, which operates on spatial autoresonance principles in a $\mathrm{TM}_{110}$ transverse magnetic mode. This acceleration mechanism has also been theoretically studied, further exploring its feasibility and optimization \cite{orozco2024electron}.

\noindent Another notable advancement is Spatial Autoresonance Acceleration (SARA), which utilizes a transverse electric field in the \(TE_{11p}\) mode (\(p = 1, 2, 3\)). This method has been theoretically studied \cite{PhySARA}, and has led to the patenting of an X-ray source based on the SARA concept \cite{dugar2017compact}. In this study, we focus on the design and numerical simulation of a RF accelerator based on the SARA mechanism. Using a cylindrical cavity excited in the $\mathrm{TE}_{112}$ mode, electron acceleration was modeled with COMSOL Multiphysics, allowing us to derive practical design parameters. Our simulation results show good agreement with those obtained via numerical solutions of the relativistic Newton-Lorentz equation using the Boris method, as reported in \cite{sara2008,sara2016i}. While the potential of the SARA mechanism for compact X-ray source development has been previously identified, this work makes a significant contribution by providing optimized design parameters for its practical implementation.

\section{Theoretical Foundation and Simulation Framework of the SARA-Based RF Accelerator}
\subsection{SARA Mechanism}
This study focuses on an RF accelerator based on the Spatial Autoresonance Acceleration (SARA) mechanism \cite{PhySARA,dugar2017compact,otero2019numerical}, designed to accelerate low-energy electron beams within a resonant cavity excited in a \(\mathrm{TE}_{11p}\) mode (\(p = 1, 2, 3\)). This mechanism enables sustained electron acceleration by keeping them in a regime close to the electron cyclotron resonance (ECR) condition along their trajectory, achieved through a carefully tailored magnetostatic field. This approach facilitates efficient and continuous energy transfer within a compact structure.

\begin{figure}[ht]
    \centering
    \includegraphics[scale=0.35]{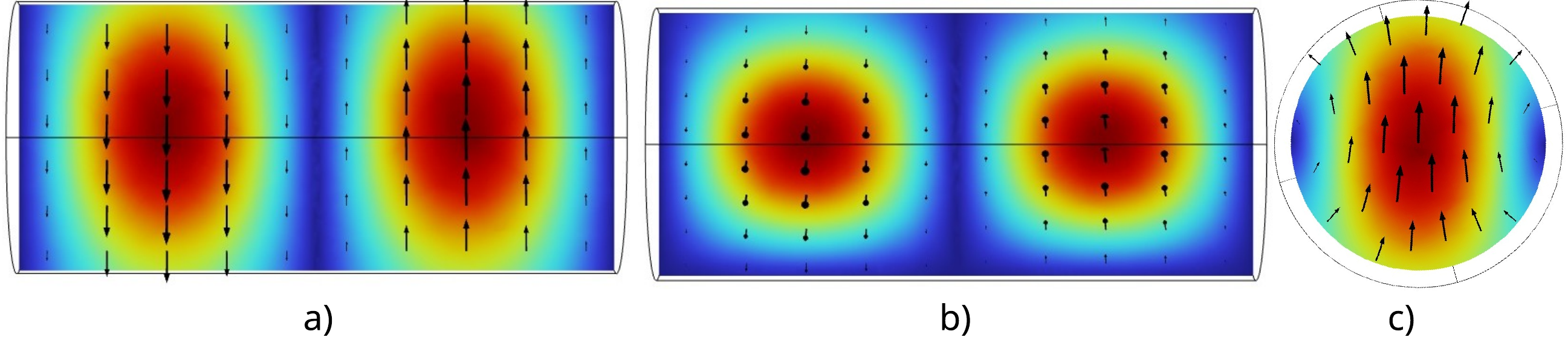}
    \caption{Electric field vector profile and corresponding field strength distribution of the \(TE_{112}\) mode in the planes: (a) \(x = 0\), (b) \(y = 0\), and (c) \(z = L/4\). These visualizations illustrate the spatial structure and intensity of the electric field components within the resonant cavity.}
    \label{fig:TE112_Mode}
\end{figure}

\noindent In a vacuum cylindrical cavity with perfect electric conductor (PEC) boundaries, the electric and magnetic field components of the \(\mathrm{TE}_{11p}\) mode with linear polarization are given by:
\begin{equation}
\begin{aligned}
E_r^{hf} & =-\frac{2 E_{0 l}}{k_{\perp} r} J_1\left(k_{\perp} r\right) \sin \theta \sin\left(k_{z} z\right) \sin{\left(\omega t+\psi_0\right)}, \\
E_\theta^{hf} & =-2 E_{0 l} J_1^{\prime}\left(k_{\perp} r\right) \cos \theta \sin\left(k_{z} z\right)  \sin{\left(\omega t+\psi_0\right)}, \\
B_r^{hf} & =2 E_{0 l} \frac{k_{z}}{\omega} J_1^{\prime}\left(k_{\perp} r\right) \cos \theta \cos \left(k_{z} z\right) \cos{\left(\omega t+\psi_0\right)}, \\
B_\theta ^{hf} & =-\frac{2 E_{0 l}}{k_{\perp} r} \frac{k_{z}}{\omega} J_1\left(k_{\perp} r\right) \operatorname{sin} \theta \cos \left(k_{z} z\right) \cos{\left(\omega t+\psi_0\right)}, \\
B_z ^{hf} & =2 E_{0 l} \frac{k_{\perp}}{\omega} J_1\left(k_{\perp} r\right) \cos \theta \operatorname{sin}\left(k_{z} z\right) \cos{\left(\omega t+\psi_0\right)}, \label{TE_components}
\end{aligned}
\end{equation} 
where the superscript ``hf" refers to the high-frequency field. The parameters \(k_\perp = S_{11} / a\) (with \(S_{11} = 1.841)\) and \(k_z = p \pi / L\)   represent the transverse and longitudinal wavenumbers, respectively. Here, \(L\) and \(a\) denote the length and radius of the cavity. The parameter \(E_{0l}\) corresponds to the maximum amplitude of the electric field, while the angular frequency of the electromagnetic wave is given by:
\begin{equation}
\omega=c \sqrt{k_z^2+k_{\perp}^2} \label{Frec_res},
\end{equation}
where $c$ is the speed of light in vacuum. Additionally, the parameter $\psi_0$ is an arbitrary phase. In this work, we focus on the \(\mathrm{TE}_{112}\) mode, whose electric field distribution  is illustrated in Fig.\ref{fig:TE112_Mode}.

\noindent When \(K_\perp r< 1\), the Bessel function \(J_1(K_\perp r)\) can be approximated as \(J_1(K_\perp r) \approx K_\perp r / 2\). Under this approximation, the field equations describe a linearly polarized standing plane wave; which can also be expressed as the superposition of two circularly polarized waves: one right-handed and one left-handed. The electric field of the microwave field is given by:
\begin{equation}
\mathbf{E}=\mathbf{E}^l+\mathbf{E}^r,
\end{equation}
where the components are:

\begin{equation} \label{E_left}
   \mathbf{E}^l =\frac{E_{0l}}{2}\sin \left(k_z z\right)[\sin (\omega t+\psi_0) \mathbf{\hat{i}}+\cos (\omega t+\psi_0) \mathbf{\hat{j}}],
\end{equation}
and
\begin{equation} \label{E_right}
\mathbf{E}^r =\frac{E_{0l}}{2}\sin \left(k_z z\right)[-\sin (\omega t+\psi_0) \mathbf{\hat{i}}+\cos (\omega t+\psi_0) \mathbf{\hat{j}}].
\end{equation}

\noindent The right-handed component plays a fundamental role in the autoresonant interaction with electrons, as it aligns with their gyration direction, enabling energy transfer.

\begin{figure}[!htb]
	\centering
	\includegraphics*[width=1.0\columnwidth]{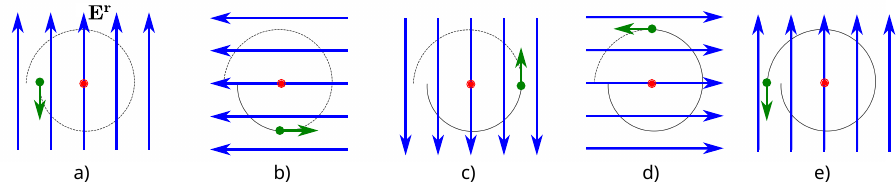}
	\caption{ Electron resonance interaction with the right-hand polarization (RHP) component at different time instants: (a) \( t=0 \), (b) \( t=T/4 \), (c) \( t=T/2 \), (d) \( t=3T/4 \), and (e) \( t=T \), where \( T = 2\pi / \omega \) represents the period of the electromagnetic wave. The blue arrows depict the right-handed electric field component of the high-frequency field, the green arrow indicates the electron velocity, and the red circle represents the magnetostatic field oriented out of the page. The dashed line illustrates the complete electron trajectory over one period of the electromagnetic wave, while the solid black line represents the electron’s path from \( t=0 \) to the indicated time instant. The Larmor radius increases as the electron gains energy.}
	\label{fig:ECR_Righ_Hand}
\end{figure}

\noindent The Spatial Autoresonance Acceleration (SARA) mechanism leverages the electron cyclotron resonance (ECR) phenomenon, in which electrons in a magnetic field efficiently absorb energy from the right-hand circularly polarized component of the wave (See Eq. \ref{E_right}). This resonance occurs when the electromagnetic wave frequency matches the electron's cyclotron frequency:
\begin{equation}
 \omega=\omega_c,
\end{equation}
where $\omega_c$ is the electron cyclotron frequency. Under this condition, the electric field remains in phase with the electron’s motion, allowing continuous energy absorption. For a right-hand circularly polarized wave, the electric field rotates in the same direction as the electron’s gyro-motion, ensuring efficient energy transfer (See Fig. \ref{fig:ECR_Righ_Hand}). 

\noindent In a uniform magnetic field, an electron follows a helical trajectory. The electron's gyro-motion occurs at the cyclotron frequency, defined as: 
\begin{equation}
\omega_c=\frac{e B}{m_e \gamma},
\end{equation}
where \(e\) is the electron charge, \(m_e\) is the electron mass, and \(B\) represents the magnetic field strength. The relativistic Lorentz factor is given by \(\gamma = \left(1 - \beta^2\right)^{-1/2}\), where \(\beta = v / c\) denotes the ratio of the electron's velocity \(v\) to the speed of light \(c\) in vacuum.

\noindent As the electron interacts with the right-hand circularly polarized (RHP) component of the microwave field (See Fig. \ref{fig:ECR_Righ_Hand}), it absorbs energy, leading to an increase in its Larmor radius while maintaining synchronization with the wave. This autoresonant interaction ensures a continuous energy transfer, sustaining the electron acceleration.

\noindent In the SARA scheme, a specially designed magnetostatic field is employed to maintain the electron cyclotron resonance (ECR) condition:
\begin{equation}
    \omega = \omega_c \approx \frac{e B_z^s(0,z)}{\gamma m_e},
\end{equation}
where $ B_z^s(0,z)$ represents the longitudinal component of the magnetostatic field along the symmetry axis ($r=0$). This component can be expressed as:
\begin{equation}
     B_z^s(0,z)=B_0[1+b(z)].
\end{equation}
where $B_0=$ $\omega m_e / e$ defines the classical resonance magnetic field, and $b(z)$ is a dimensionless function that varies along the cavity axis. The function $b(z)$ accounts for the necessary adjustments required to maintain resonance with the electromagnetic wave at frequency $\omega$, ensuring efficient energy transfer to the electrons.

\noindent For the $\mathrm{TE}_{111}$ mode, the magnetostatic field increases monotonically. However, in the general case of the $\mathrm{TE}_{11p}$ mode with $p \neq 1$, the magnetostatic field exhibits a nonmonotonic behavior, requiring careful adjustments to maintain the phase shift between the electron velocity and the electric field, $\varphi$, within the range:
\begin{equation}
    \pi / 2 < \varphi < 3\pi / 2.
\end{equation}
This interval, referred to as the ``Acceleration Band", ensures that the electromagnetic field efficiently transfers energy to the electrons.  

\noindent To achieve this in the SARA scheme, for this type of microwave modes, the magnetostatic field must locally decrease in the region immediately following each microwave field node and then gradually increase until reaching the next node \cite{orozco2019simulation}. This tailored magnetic field profile is crucial for maintaining autoresonance and ensuring continuous electron acceleration.

\noindent In spatial autoresonance acceleration, the longitudinal component of the magnetic force acts as a diamagnetic force, opposing the magnetic field gradient and restricting the electrons' motion toward regions of higher magnetic field strength. As a result, the acceleration process terminates when the longitudinal velocity of the electrons vanishes due to the cumulative effect of this diamagnetic force.

\subsection{Physical scheme: The SARA-Based RF accelerator}

\noindent Figure \ref{fig:Physical_Scheme} illustrates an RF accelerator based on the cylindrical $\mathrm{TE}_{112}$ mode. The system operates by utilizing a resonant cavity and a tailored magnetostatic field to sustain the spatial autoresonance acceleration (SARA) mechanism, ensuring continuous energy transfer to the electrons.

\subsubsection*{Microwave Injection and Excitation}
\noindent Microwaves are introduced into the system from a source (not shown in the figure) into a rectangular waveguide (1), where they propagate in the $\mathrm{TE}_{10}$ mode. These waves are subsequently coupled into the cylindrical resonant cavity (3) via a tapered waveguide (2), efficiently exciting the $\mathrm{TE}_{112}$ mode within the cavity. This particular mode exhibits a field distribution that closely approximates a standing plane wave, making it highly suitable for efficient electron acceleration.

\subsubsection*{Resonant Cavity and Electromagnetic Field Configuration}
\noindent The resonant cavity is modeled as a copper structure, a material with high electrical conductivity and non-magnetic properties. Its high conductivity reduces ohmic losses, enhancing overall efficiency. Additionally, due to its non-magnetic nature, the cavity remains permeable to the applied magnetostatic field, allowing it to influence the electron dynamics without distortion. The electromagnetic field inside the cavity is carefully designed to sustain autoresonance, ensuring that electrons remain in phase with the right-handed circularly polarized component of the wave for continuous acceleration.

\subsubsection*{Magnetostatic Field Generation and Electron Acceleration}
\begin{figure}[ht]
    \centering
    \includegraphics[scale=0.6]{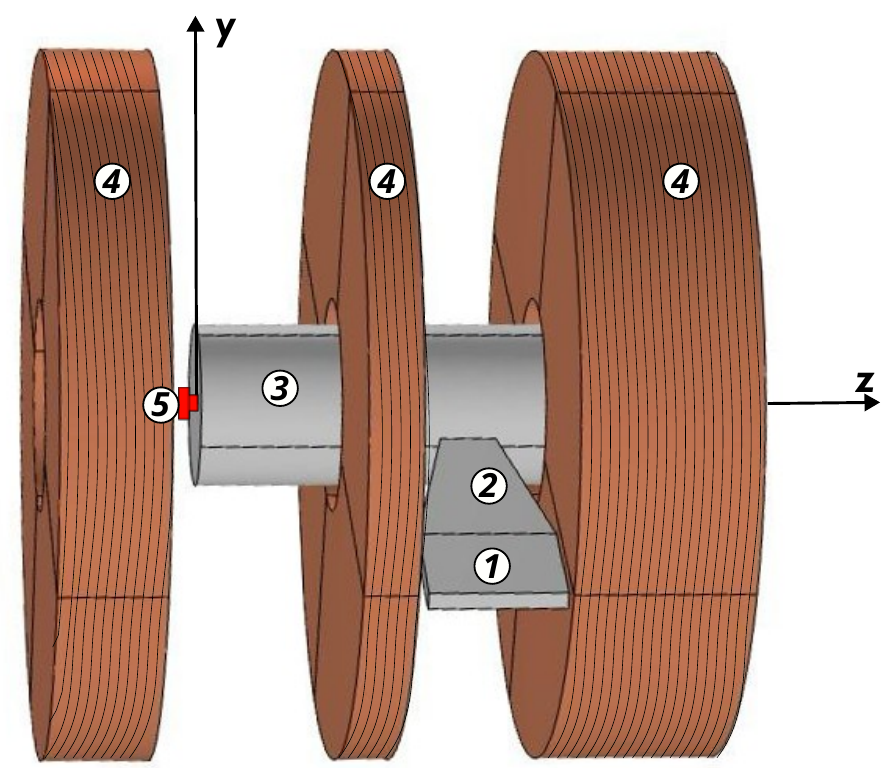}
    \caption{Schematic representation of the SARA-based RF accelerator. Microwaves are injected into a rectangular waveguide (1), where they propagate in the $\mathrm{TE}_{10}$ mode. The wave is then coupled into a cylindrical resonant cavity (3) through a tapered waveguide (2), exciting the $\mathrm{TE}_{112}$ mode. The cavity is influenced by a tailored magnetostatic field generated by a set of direct current (DC) coils (4), designed to maintain the electron cyclotron resonance (ECR) condition along the electrons path. Electrons are injected on-axis by an electron gun (5) and undergo sustained acceleration due to the combined effects of the RF and magnetostatic fields. Upon reaching high energies, the electrons impact a metallic target positioned at the opposite end of the cavity (not shown), generating X-ray radiation via Bremsstrahlung emission.}
    \label{fig:Physical_Scheme}
\end{figure}

\noindent The required magnetostatic field is generated by a set of direct current (DC) coils (4), as depicted in Figure \ref{fig:Physical_Scheme}. This nonmonotonic magnetic field is carefully designed to maintain the electron cyclotron resonance (ECR) condition along the cavity axis, optimizing energy transfer from the microwave field to the electrons. These electrons are injected along the cavity axis using an electron gun (5), with an injection energy determined by the configured microwave and magnetostatic fields. The superposition of the RF field of the $\mathrm{TE}_{112}$ mode and the tailored magnetostatic field provides the conditions for sustained electron acceleration. Under optimal conditions, electrons can reach energies of up to 200 keV. These high-energy electrons, upon impacting a metallic target, generate X-ray radiation via Bremsstrahlung emission, making the system suitable for applications in medical imaging, security scanning, and materials analysis.

\subsubsection*{System Efficiency and Compact Design}
\noindent The integration of the RF field with the precisely tailored magnetostatic field results in a highly compact and efficient accelerator design. This approach minimizes system footprint while maximizing energy transfer efficiency, demonstrating strong potential for experimental implementation and practical applications requiring compact electron accelerators.

\subsection{Simulation Methodology}

\begin{figure}[ht]
    \centering
    \includegraphics[scale=0.5]{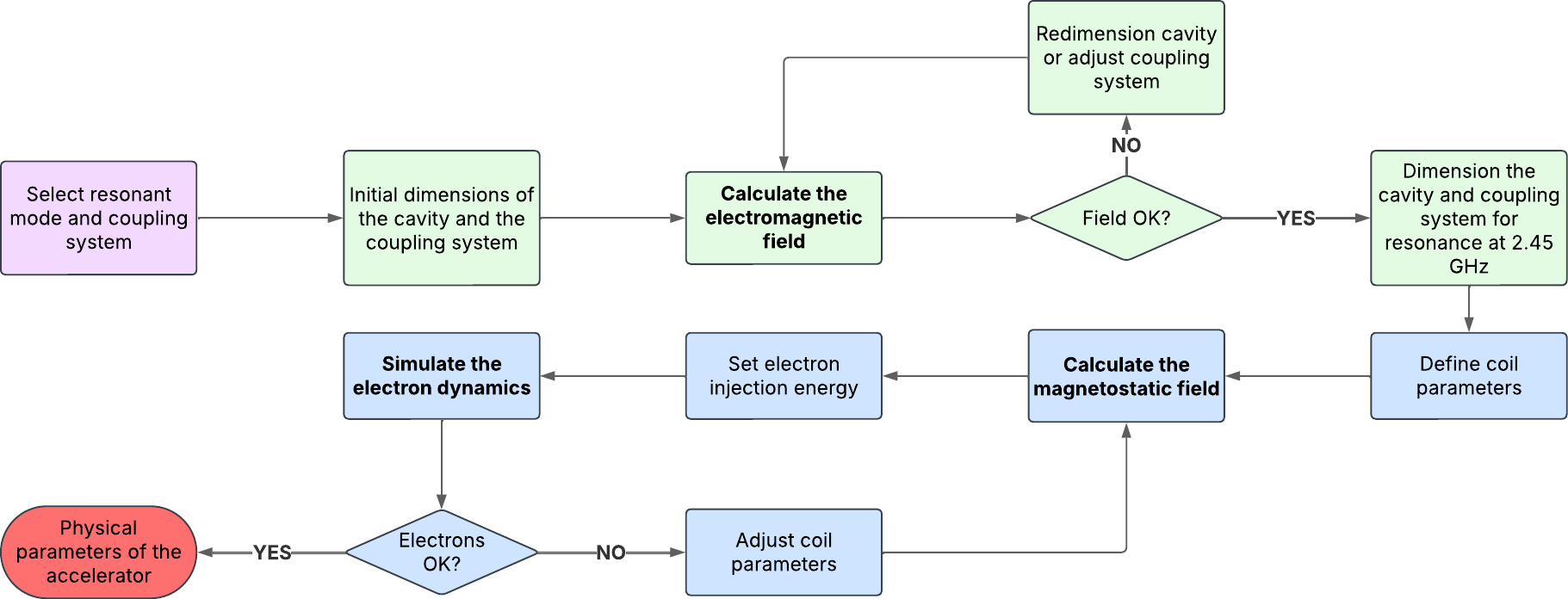}
    \caption{Flowchart illustrating the methodology used for the design and optimization of the SARA-based RF accelerator. The process involves iterative simulations and parameter adjustments to achieve the desired electromagnetic and magnetostatic field configurations, ensuring efficient electron acceleration.}
    \label{fig:Methodology}
\end{figure}

For the design of the SARA-based RF accelerator, we followed the methodology illustrated in Fig. \ref{fig:Methodology}. The appropriate electromagnetic fields for developing a SARA-based RF accelerator are primarily based on the transverse electric \(\mathrm{TE}_{11 p}\) cylindrical modes. In the present case, the \(\mathrm{TE}_{112}\) mode was chosen as the resonant mode. After selecting the resonant mode and the coupling system, the cavity dimensions were determined through electromagnetic field simulations to achieve resonance at 2.45 GHz. The intensity and spatial distribution of the electromagnetic field were then analyzed. If the results were not satisfactory, the cavity was resized, or adjustments were made to the coupling system parameters using a parametric sweep. This process was repeated until an appropriate field configuration was achieved, ensuring that the chosen mode was excited at the desired frequency \((2.45 \mathrm{GHz})\) and that the electric field reached the required intensity.

\noindent Once the electromagnetic field was  established, the coil parameters (including dimensions, axial positions, and currents) were defined to calculate the magnetostatic field. The initial conditions for electron injection were then specified, and the dynamics of a single electron were simulated under the influence of the calculated electromagnetic field. If the electron impacted the cavity's rear wall with insufficient or excessive energy, the coil parameters were adjusted, and the simulation was repeated. This iterative process continued until a satisfactory magnetostatic field configuration was achieved.

\noindent This methodology remains applicable for the design of similar RF accelerators and can be adapted for further optimizations.

\noindent The RF accelerator based on the SARA mechanism was simulated using COMSOL Multiphysics, a widely used commercial software for realistic modeling of complex multiphysics systems \cite{pryor2009multiphysics,zimmerman2006multiphysics}. In this study, the Radiofrequency (RF) and AC/DC modules of COMSOL Multiphysics were employed to accurately model the electromagnetic fields, magnetostatic configuration, and electron dynamics within the accelerator.

\noindent In this simulation framework, the electromagnetic waves are solved in the frequency domain. The formulation begins with Maxwell’s equations in the frequency domain \cite{schwarzbach2009stability}:

\begin{equation}
 \nabla \times \widehat{\boldsymbol{E}}^{hf}(\boldsymbol{r}, \omega) = -\mathrm{j} \omega \widehat{\mu}(\omega) \widehat{\boldsymbol{H}}^{hf}(\boldsymbol{r}, \omega),   
\end{equation}
and
\begin{equation}
\nabla \times \widehat{\boldsymbol{H}}^{hf}(\boldsymbol{r}, \omega)=\widehat{J}(\boldsymbol{r}, \omega)+\mathrm{j} \omega \widehat{\boldsymbol{D}}^{hf}(\boldsymbol{r}, \omega),
\end{equation}
where the hat operator ($\text{ } \widehat{} \text{ }$) represents the Fourier transform of the corresponding physical quantities. Here, \(\widehat{\boldsymbol{E}}^{hf}(\boldsymbol{r}, \omega)\) and \(\widehat{\boldsymbol{H}}^{hf}(\boldsymbol{r}, \omega)\) denote the electric field and magnetic field intensity at the frequency $\omega$, respectively, while \(\widehat{\mu}(\omega)\) is the magnetic permeability.

\noindent The current density is given by:
\begin{equation}
  \widehat{J}(\boldsymbol{r}, \omega)=\widehat{\sigma}(\omega) \widehat{\boldsymbol{E}}^{hf}(\boldsymbol{r}, \omega),
\end{equation}
where $\widehat{\sigma}(\omega)$ is the electrical conductivity of the medium. The electric displacement field is given as:
\begin{equation}
 \widehat{\boldsymbol{D}}^{hf}(\boldsymbol{r}, \omega) = \varepsilon_0 \widehat{\varepsilon}_r(\omega) \widehat{\boldsymbol{E}}^{hf}(\boldsymbol{r}, \omega),
\end{equation}
 where \(\varepsilon_0\) is the vacuum permittivity and \(\widehat{\varepsilon}_r(\omega)\) is the relative permittivity of the medium.

\noindent By combining these equations, the electric field component of the electromagnetic wave is obtained as:

\begin{equation}
\nabla \times \widehat{\mu}_r(\omega)^{-1} \nabla \times \widehat{\boldsymbol{E}}^{hf}(\boldsymbol{r}, \omega)-\omega^2 \varepsilon_0 \mu_0\left(\widehat{\varepsilon}_r(\omega)-\frac{\mathrm{j} \hat{\sigma}(\omega)}{\omega \varepsilon_0} \right) \widehat{\boldsymbol{E}}^{hf}(\boldsymbol{r}, \omega)=0. \label{wave_equation}
\end{equation}

\noindent For the simulation of the microwave field inside the cavity,  the following parameters are used in Eq. (\ref{wave_equation}):
\begin{equation}
\hat{\varepsilon}_r(\omega)=1, \quad \widehat{\mu}_r(\omega)=1, \quad \text { and } \quad \widehat{\sigma}(\omega)=0 .
\end{equation}

\noindent These values are chosen because the resonant cavity operates in a vacuum. The assumption $\sigma=0$ remains valid even in the presence of non-ionized residual gases.

\noindent To characterize the electromagnetic behavior of the system, two key analyses were performed. First, a natural frequency study was conducted to determine the resonance frequencies and the corresponding natural modes of the cavity. This analysis identified the optimal excitation frequency for establishing the desired mode and guided the selection of an appropriate coupling mechanism. Subsequently, a frequency domain study was carried out to evaluate the response of a linear or linearized system subjected to harmonic excitation.  The cylindrical cavity was then excited at the resonance frequency using a tapered waveguide. A parametric sweep of the waveguide's geometric parameters was performed to optimize the excitation efficiency of the desired mode within the cavity 

\noindent The AC/DC module was employed to compute the magnetostatic field generated by the coil system carrying currents (see Figure \ref{fig:Physical_Scheme}) and to analyze the acceleration of injected electrons along the axis of the resonant cavity.

\noindent The governing equations for the magnetostatic field in this setup are given by:

\begin{equation}
\begin{aligned}
& \nabla \times \mathbf{H}^s=\mathbf{J}, \\
& \mathbf{B}^s=\mu_o \mathbf{H}^s, \\
& \mathbf{J}=\sigma \mathbf{E}+\mathbf{J}_{\mathrm{e}},
\end{aligned}
\end{equation}
where \(\mathbf{H}^s\) represents the magnetic field intensity, \(\mathbf{B}^s\) denotes the magnetic flux density, \(\sigma\) is the electrical conductivity of the medium, and, \(\mathbf{J}_{\mathrm{e}}\) corresponds to the external current density.

\noindent As in the previous analysis, the electrical conductivity was set to zero $(\sigma=0)$ for this study.

\noindent The current density of the coils is determined using the following expression:
\[
\mathbf{J}_{\mathrm{e}} = \frac{N I_{\mathrm{coil}}}{A_c},
\]
\noindent where $N$ is the number of turns in the coil, $I_{\text {coil }}$ is the current flowing through the coil, $A_c$ is the cross-sectional area of the coils.

\noindent To ensure continuous electron acceleration, the magnetostatic field must follow a specific profile to maintain the resonance condition. The optimization of the field configuration was performed using a parametric sweep, taking previous studies as an initial reference \cite{sara2016}.

\noindent The motion of electrons in the RF accelerator is modeled using the single-particle approximation, governed by the relativistic Newton-Lorentz equation:

\begin{equation}
\frac{d\left(\gamma m_e \boldsymbol{v}\right)}{d t}=-e[\boldsymbol{E}+(\boldsymbol{v} \times \boldsymbol{B})],
\label{Newton_Lorentz}
\end{equation}

\noindent where \(\gamma = \left[1 - (v / c)^2\right]^{-1/2}\) is the Lorentz factor, \(m_e\) and \(-e\) are the mass and charge of the electron, and \(\boldsymbol{v}\) is its velocity.  The fields governing the electron's motion are given by \(\boldsymbol{E} = \boldsymbol{E}^{hf}\) and \(\boldsymbol{B} = \boldsymbol{B}^{hf} + \boldsymbol{B}^s\), where \(\boldsymbol{E}^{hf}\) and \(\boldsymbol{B}^{hf}\) correspond to the electric and magnetic components of the high-frequency electromagnetic field excited in the \(\mathrm{TE}_{112}\) mode in the cavity. \(\boldsymbol{B}^s\) represents the axisymmetric static magnetic field.
\section{Results and Discussion}
\subsection{The electromagnetic simulation}
\noindent In this study, a resonant cavity made of copper was considered, with an electrical conductivity of \(5.96\times~10^7\,~\mathrm{S/m}\). Using the \textit{emw} interface of the RF module in COMSOL Multiphysics, a natural frequency study was performed to determine the optimal cavity dimensions required to excite the \(\mathrm{TE}_{112}\) mode at a frequency of 2.45 GHz. Initially, a perfectly cylindrical cavity (without microwave injection ports) was analyzed.

\begin{figure}[ht]
    \centering
    \includegraphics[scale=0.35]{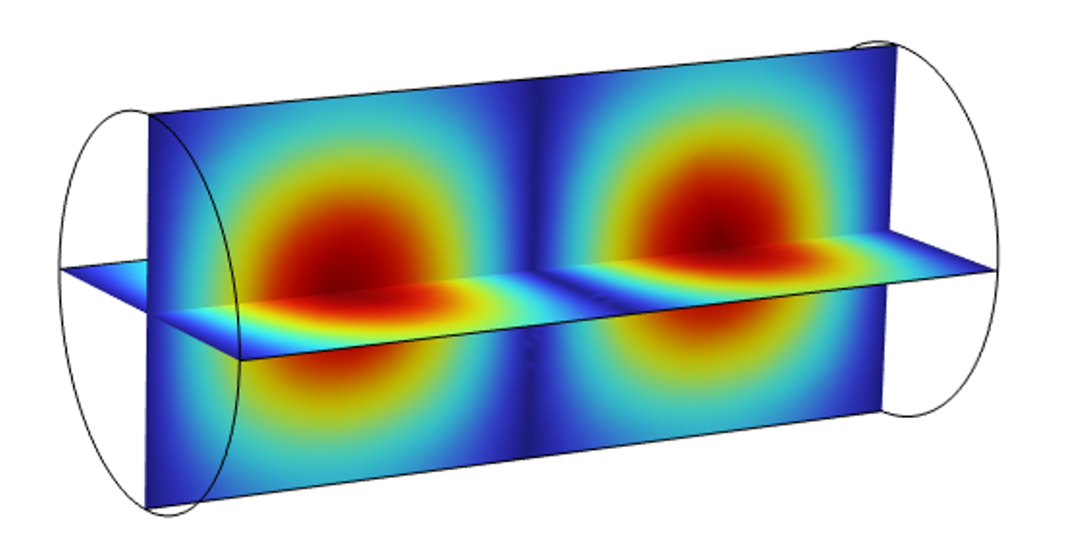}
    \caption{Electric field distribution of the pure $\mathrm{TE}_{112}$ mode in a cylindrical copper cavity, obtained through a parametric sweep. The optimized cavity dimensions for a resonance frequency of 2.45 GHz are a radius of 4.52 cm and a length of 19.8 cm.}
    \label{C245frecpro}
\end{figure}

\noindent The results indicate that a cavity with a radius of 4.52 cm and a length of 19.8 cm successfully supports the desired mode at this frequency. These dimensions are very close to those reported in \cite{sara2016i}, where the cavity walls were assumed to be perfect electrical conductors to analyze the electron acceleration via the SARA mechanism for intense electron beams using the particle-in-cell method. Figure \ref{C245frecpro} presents the corresponding electric field distribution obtained from the electromagnetic simulations.

\noindent Using the previously determined cavity radii, 4.52cm, the \(\mathrm{TE}_{112}\) mode was excited via a tapered rectangular waveguide, laterally coupled to the cavity at the position \(3L / 4\), where \(L = 20 \, \mathrm{cm}\) represents the cavity's length. A microwave power of 1 kW was considered. 

\noindent A standard WR340 rectangular waveguide was used as the input port, with internal dimensions of \(8.636 \, \mathrm{cm} \times 4.318 \, \mathrm{cm}\) (width \(\times\) height) and a length of \(6.00 \, \mathrm{cm}\). This waveguide was coupled to a tapered rectangular waveguide designed to gradually reduce the width down to \(1.89 \, \mathrm{cm}\), corresponding to the connection interface with the resonant cavity. The tapered section has a total length of \(6.00 \, \mathrm{cm}\), and a constant height of \(0.6 \, \mathrm{cm}\) was maintained throughout the transition.

\begin{figure}[ht]
    \centering
    \includegraphics[scale=1.0]{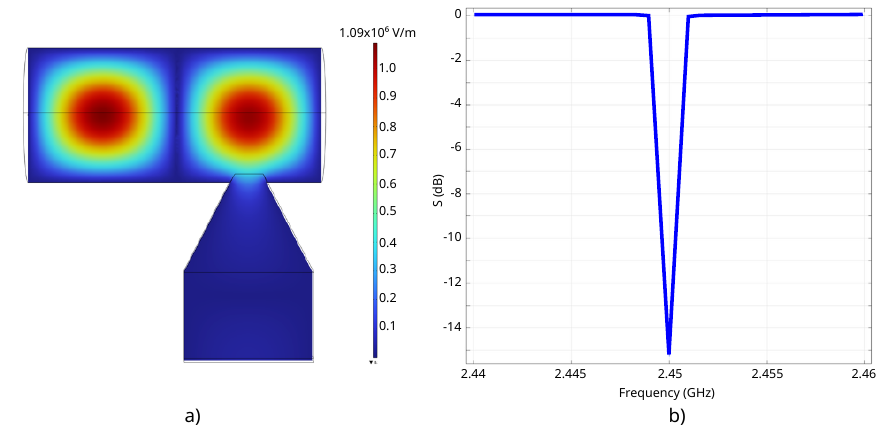}
    \caption{Electric field magnitude distribution inside the cavity at a frequency of 2.45 GHz for an input microwave power of 1 kW. The system employs a tapered rectangular waveguide with an input port measuring $8 \times 0.6$ cm (width $\times$ height) and an output port measuring $4.4 \times 0.6$ cm. The tapered section has a total length of 12.15 cm.}
    \label{C245Guiatapered}
\end{figure}

\noindent As shown in Fig. \ref{C245Guiatapered}, the resulting electric field successfully reproduces the pure \(\mathrm{TE}_{112}\) mode. For a microwave power of 1 kW, the maximum electric field intensity reached approximately $1 \times 10^6$ V/m.

\begin{figure}[ht]
    \centering
    \includegraphics[scale=0.45]{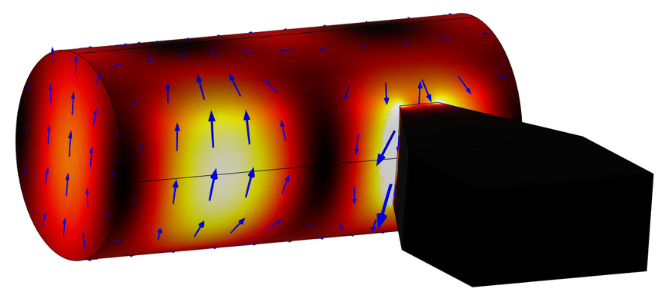}
    \caption{Ohmic losses resulting from surface currents induced on the interior walls of the cavity. The color scale indicates the local power dissipation, with peak values reaching approximately \(5 \times 10^{-3} \, \mathrm{kW/cm^2}\).}
    \label{Ohmic_losses}
\end{figure}

\noindent Figure \ref{Ohmic_losses} illustrates the Ohmic losses caused by surface currents induced on the interior walls of the cavity, emphasizing the correlation between the electric field distribution (See Fig. \ref{C245Guiatapered}) and the intensity of these losses. The maximum dissipated power is approximately \(5\times10^{-3} \, \mathrm{kW/cm^2}\). For the selected microwave coupling system, a quality factor of \(Q \sim 25000\) was obtained, confirming its effectiveness in minimizing energy dissipation and ensuring high efficiency in electromagnetic energy storage.

\noindent To efficiently excite the electromagnetic field in the high-Q resonant cavity, a microwave generator with precise frequency control and high stability is required. Given that the cavity exhibits a quality factor of $Q\sim25000$, the corresponding resonance bandwidth is approximately $100$ kHz, which imposes stringent requirements on the microwave source. The frequency stability of the generator must be well within this bandwidth to ensure continuous operation at the resonant frequency ($2.45$ GHz) without significant detuning.

\noindent Since the resonance curve is relatively narrow, even small frequency drifts could cause the system to move out of resonance, reducing the efficiency of energy transfer and electron acceleration. To mitigate this challenge, the microwave source should meet the following criteria: 
\begin{itemize}
    \item[1] \textbf{Fine frequency resolution} (ideally $\leq 1$ kHz) to allow precise tuning within the resonance curve.
    \item[2] \textbf{High frequency stability} (preferably $\pm 10$ ppm or better) to minimize frequency fluctuations that could shift the system out of resonance.
\end{itemize}

\noindent Considering the high-Q nature of the cavity, selecting a microwave generator with a well-controlled phase noise profile and stability-enhancing features would significantly improve the efficiency of energy transfer, ensuring optimal excitation of the electromagnetic field and sustained electron acceleration.

\noindent Additionally, temperature control is recommended to minimize thermal expansion of the cavity walls, which could lead to frequency shifts affecting $\omega_0$. Furthermore, if the bandwidth of the microwave source does not fully match the required resonance bandwidth ($\Delta \omega$), alternative strategies such as adjustable coupling mechanisms or slight modifications to the cavity geometry may be considered to balance $Q$ and tuning flexibility.
  
\subsection{The magnetostatic field simulation}
In the second phase, the \textit{mf} interface of the \textit{AC/DC} module in COMSOL was used to determine the magnetostatic field profile required to maintain the resonance condition. Starting with the coil parameters presented in \cite{sara2008}, a parametric sweep was performed by varying the current in three coils, computing the resulting magnetic field, and simulating the dynamics of a single electron under the previously calculated microwave field and the corresponding magnetostatic field, as described in the methodology illustrated in Fig. \ref{fig:Methodology}. 

\begin{figure}[ht]
    \centering
    \includegraphics[scale=1.2]{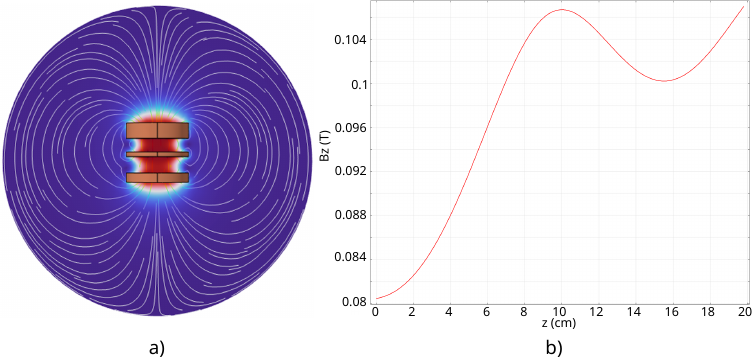}
    \caption{a) Magnetic field lines of the three-coil system used to generate the required magnetostatic field. b) Magnetostatic field profile along the cavity axis. The coil currents were optimized to maintain the resonance condition along the cavity axis for the power level of 1kW.}
    \label{C245tBz}
\end{figure}

\noindent Figure \ref{C245tBz}a depicts the magnetic field lines generated by the three-coil system, while Figure \ref{C245tBz}b presents the magnetostatic field profile along the cavity axis, which sustains the electron resonance condition along its trajectory.

\begin{table}[ht]
\centering
\begin{tabular}{|c|c|c|c|c|c|c|}
\hline
\multicolumn{5}{|c|}{} & \multicolumn{2}{c|}{\textbf{1 kW}} \\
\hline
\textbf{Coil} & $r_i(cm)$ & $r_e(cm)$ & $z_c(cm)$ & $\Delta z(cm)$ & $I (A)$ & \begin{tabular}{c}
$J$ \\
$\left(\frac{mA}{\text{mm}^2}\right)$
\end{tabular}\\
\hline
1 & 6 & 20 &-5.75 & 6 & 11.11 & 1.32 \\
\hline
2 & 6 & 20 & 9.25 & 2 & 12.70 & 4.54 \\
\hline
3 & 6 & 20 & 22.5 & 8 & 17.80 & 1.59 \\
\hline
\end{tabular}
\caption{Geometric and electrical parameters of the three-coil system used to generate the required magnetostatic field. The table lists the axial positions (\(z_c\)), coil dimensions (\(r_i\), \(r_e\), \(\Delta z\)), where \(r_i\) and \(r_e\) represent the inner and outer coil radii, respectively, and \(\Delta z\) denotes the coil width. Additionally, it includes the total current values at the coils.}
\label{tab:coil_parameters}
\end{table}

\noindent For the electron simulations, the following conditions were considered: 
\begin{itemize}
    \item \textbf{Injection point:} \(z = 0\) and \(r = 0\) 
    \item \textbf{Injection energy:} $E = 8, 10, 12, 14 \, \mathrm{keV}$
    \item \textbf{Time step:} \(\Delta t = \frac{2\pi}{50 \omega_c}\)  
\end{itemize}

\begin{figure}[ht]
    \centering
    \includegraphics[scale=0.45]{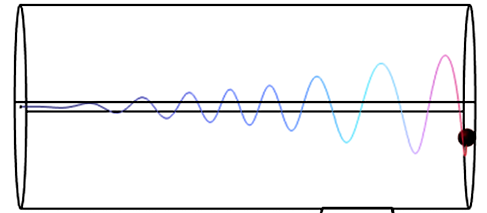}
    \caption{Simulated spiral trajectory of an electron projected onto the longitudinal plane. The electron is injected along the cavity axis (\(r_0 = 0\)) with an initial energy of 10 keV. As it interacts with the electromagnetic fields, its Larmor radius increases, indicating continuous energy gain until it impacts the opposite wall of the cavity.}
    \label{Trajectory}
\end{figure}

\noindent Figure \ref{Trajectory} shows the computed spiral trajectory of an electron projected onto the longitudinal plane. The electron is injected along the cavity axis (\(r_0=0\)) with an initial energy of 10 keV. The progressive increase in the Larmor radius indicates a continuous energy gain as the electron interacts with the electromagnetic field, until it eventually impacts the opposite wall of the cavity.  

\begin{figure}[ht]
    \centering
    \includegraphics[scale=0.9]{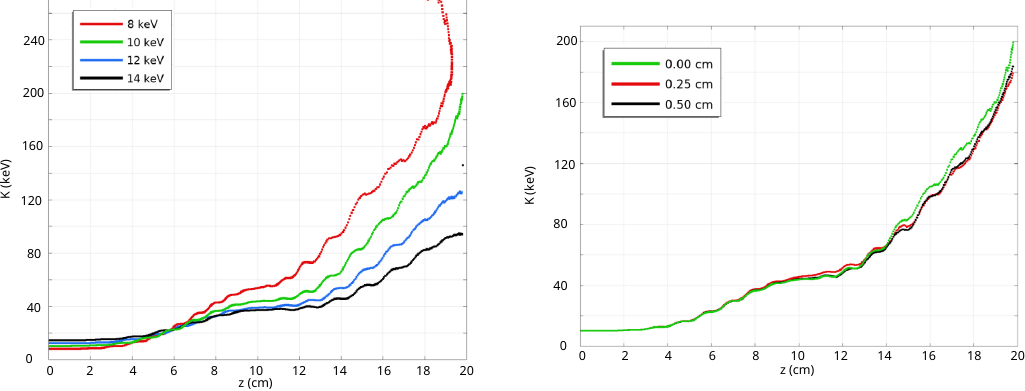}
        \caption{(a) Spatial evolution of the electron energy for different injection energies (\(8, 10, 12\), and \(14 \, \mathrm{keV}\)) at a fixed microwave power of \(1 \, \mathrm{kW}\). Although the 8 keV case yields the highest energy gain, the electron is reflected before reaching the cavity exit due to the diamagnetic effect. (b) Spatial evolution of the electron energy for different radial injection positions, showing improved resonance when the electron is injected along the cavity axis (\(r = 0 \, \mathrm{cm}\)).}
    \label{C245E0}
\end{figure}

\noindent Figure~\ref{C245E0} presents the spatial evolution of electron energy for different injection energies (\(8, 10, 12\), and \(14 \, \mathrm{keV}\)) and for different radial injection positions, all at a microwave power level of \(1 \, \mathrm{kW}\). In Figure~\ref{C245E0}(a), the highest energy gain corresponds to an injection energy of \(8 \, \mathrm{keV}\); however, it can be observed that in this case the electron tends to reverse its motion before reaching the end of the cavity. This behavior is attributed to the inhomogeneity of the magnetostatic field and is a well-known phenomenon referred to as the diamagnetic effect. In contrast, for the other injection energies, this effect is not present, leading to improved stability and control. Among the cases analyzed, an injection energy of \(10 \, \mathrm{keV}\) appears to provide the most favorable scenario.

\noindent For higher injection energies (\(12 \, \mathrm{keV}\) and \(14 \, \mathrm{keV}\)), the maximum energy attained upon reaching the cavity exit decreases. This effect is due to the reduced interaction time between the electron and the microwave field as the initial velocity increases, thereby limiting the total energy transfer.

\noindent Figure~\ref{C245E0}(b) shows the effect of injecting the electron at different radial positions, highlighting a more effective resonant interaction when the injection occurs at \(r = 0 \, \mathrm{cm}\).

\noindent Additionally, in the vicinity of \(z = 10cm\) cm, the energy remains nearly constant. This occurs because the electric field of the TE$_{112}$ mode has a node at \(L/2 = 10\) cm (see Fig. \ref{C245Guiatapered}). The energy fluctuations observed in both cases can be attributed to the interaction of the electron with the left-hand circular polarization component of the TE$_{112}$ microwave field, as previously described using an analytical approach (see Eq. (\ref{E_left})).

\section{Conclusions}
In this paper, we present the design and analysis of an RF accelerator based on the Spatial Autoresonance Acceleration (SARA) mechanism, utilizing the cylindrical TE$_{112}$ mode. The study was conducted using the Radio Frequency (RF) and AC/DC modules of COMSOL Multiphysics. A systematic approach was implemented to simulate a complex three-dimensional system and assess the interdependence of key technological parameters. The main conclusions of this work are as follows:

\noindent The results indicate that in the SARA-based RF accelerator, an electron beam can be accelerated to energies close to 200 keV using a 1 kW microwave generator. These electrons possess sufficient energy to generate X-rays upon impact with a metallic target. This highlights the system's potential for applications requiring compact X-ray sources.

\noindent This work serves as a methodological guide for the design of SARA-based RF accelerator. The results obtained are valuable both for understanding the physical principles underlying the accelerator and for establishing design parameters for its experimental implementation. The proposed system demonstrates significant potential for applications requiring compact and efficient X-ray sources, such as medical imaging, airport security, and materials analysis, reinforcing its versatility and practical relevance.

\noindent The high quality factor ($Q\sim25000$) of the resonant cavity ensures efficient electromagnetic energy storage but also imposes stringent requirements on frequency stability and tuning. A microwave source with high frequency resolution ($\leq 1$ kHz), stability ($\pm 10$ ppm or better), and active frequency-locking mechanisms is recommended to maintain resonance conditions and optimize energy transfer. Additionally, controlling the temperature of the cavity is crucial to minimize thermal expansion, which could shift the resonance frequency and affect system performance. Implementing a cooling system would help stabilize the cavity and maintain consistent operating conditions.

\noindent The proposed physical system can be modified to incorporate a target for X-ray generation by slightly adjusting the geometry of the opposite side of the cavity. Furthermore, if the bandwidth of the microwave source does not align precisely with the required resonance bandwidth, adjustable coupling strategies or slight modifications to the cavity geometry could be explored to achieve a balance between the quality factor ($Q$) and ease of tuning.

\section*{Acknowledgments}
The authors thanks Departamento Administrativo de Ciencia, Tecnología e Innovación, Colombia, and Ministerio de Ciencia Tecnología e Innovación, Colombia, for support of this work through the announcement No. 852-2019 (project ID: 1102-852-71985) and the Universidad Industrial de Santander (UIS), Colombia, (Project ID: 9482-2665).

\bibliographystyle{unsrt}
\bibliography{Bibliografia}

\end{document}